\documentclass[12pt]{article}
\usepackage{a4wide}
\usepackage{amssymb}
\usepackage{graphicx}
\begin{document}
{\renewcommand{\thefootnote}{\fnsymbol{footnote}}
\hfill  AEI--2005--025\\ 
\medskip
\hfill gr--qc/0505057\\
\medskip
\begin{center}
{\LARGE  Elements of Loop Quantum Cosmology}\\
\vspace{1.5em}
Martin Bojowald\footnote{e-mail address: {\tt mabo@aei.mpg.de}}
\\
\vspace{0.5em}
Max-Planck-Institut f\"ur Gravitationsphysik, Albert-Einstein-Institut,\\
Am M\"uhlenberg 1, D-14476 Potsdam, Germany
\vspace{1.5em}
\end{center}
}

\setcounter{footnote}{0}

\newcommand{\lP}{\ell_{\rm P}}
\newcommand{\md}{{\rm d}}
\newcommand{\sgn}{{\rm sgn}}

\begin{abstract}
  The expansion of our universe, when followed backward in time,
  implies that it emerged from a phase of huge density, the big bang.
  These stages are so extreme that classical general relativity
  combined with matter theories is not able to describe them properly,
  and one has to refer to quantum gravity. A complete quantization of
  gravity has not yet been developed, but there are many results about
  key properties to be expected. When applied to cosmology, a
  consistent picture of the early universe arises which is free of the
  classical pathologies and has implications for the generation of
  structure which are potentially observable in the near future.
\end{abstract}

\section{Introduction} 

General relativity provides us with an extremely successful
description of the structure of our universe on large scales, with
many confirmations by macroscopic experiments and so far no conflict
with observations.  The resulting picture, when applied to early
stages of cosmology, suggests that the universe had a beginning a
finite time ago, at a point where space, matter, and also time itself
were created. Thus, it does not even make sense to ask what was there
before since ``before'' does not exist at all. At very early stages,
space was small such that there were huge energy densities to be
diluted in the later expansion of the universe that is still
experienced today. In order to explain also the structure that we see
in the form of galaxies in the correct statistical distribution, the
universe not only needs to expand but do so in an accelerated manner,
a so-called inflationary period, in its early stages. With this
additional input, usually by introducing inflation with exponential
acceleration \cite{Guth,NewInfl,InflAS} lasting long enough to expand
the scale factor $a(t)$, the radius of the universe at a given time
$t$, by a ratio $a_{\rm final}/a_{\rm initial}>e^{60}$. The resulting
seeds for structure after the inflationary phase can be observed in
the anisotropy spectrum of the cosmic microwave background (most
recently of the WMAP satellite \cite{WMAPParam}), which agrees well
with theoretical predictions over a large range of scales.

Nonetheless, there are problems remaining with the overall picture.
The beginning was extremely violent with conditions such as diverging
energy densities and tidal forces under which no theory can prevail.
This is also true for general relativity itself which led to this
conclusion in the first place: according to the singularity theorems
any solution to general relativity, under reasonable conditions on the
form of matter, must have a singularity in the past or
future \cite{SingTheo}. There, space degenerates, e.g.\ to a single
point in cosmology, and energy densities and tidal forces diverge.
From the observed expansion of our current universe one can conclude
that according to general relativity there must have been such a
singularity in the past (which does not rule out further possible
singularities in the future). This is exactly what is usually referred
to as the ``beginning'' of the universe, but from the discussion it is
clear that the singularity does not so much present a boundary to the
universe as a boundary to the classical theory: The theory predicts
conditions under which it has to break down and is thus incomplete.
Here it is important that the singularity in fact lies only a finite
time in the past rather than an infinite distance away, which could be
acceptable. A definitive conclusion about a possible beginning can
therefore be reached only if a more complete theory is found which is
able to describe these very early stages meaningfully.

Physically, one can understand the inevitable presence of
singularities in general relativity by the characteristic property of
classical gravitation being always attractive. In the backward
evolution in which the universe contracts, there is, once matter has
collapsed to a certain size, simply no repulsive force strong enough
to prevent the total collapse into a singularity. A similar behavior
happens when not all the matter in the universe but only that in a
given region collapses to a small size, leading to the formation of
black holes which also are singular.

This is the main problem which has to be resolved before one can call
our picture of the universe complete. Moreover, there are other
problematic issues in what we described so far. Inflation has to be
introduced into the picture, which currently is done by assuming a
special field, the inflaton, in addition to the matter we know. In
contrast to other matter, its properties must be very exotic so as to
ensure accelerated expansion which with Einstein's equations is
possible only if there is negative pressure. This is achieved by
choosing a special potential and initial conditions for the inflaton,
but there is no fundamental explanation of the nature of the inflaton
and its properties. Finally, there are some details in the anisotropy
spectrum which are hard to bring in agreement with theoretical models.
In particular, there seems to be less structure on large scales than
expected, referred to as a loss of power.

\section{Classical Cosmology}

In classical cosmology one usually assumes space to be homogeneous and
isotropic, which is an excellent approximation on large scales today.
The metric of space is then solely determined by the scale factor
$a(t)$ which gives the size of the universe at any given time $t$. The
function $a(t)$ describes the expansion or contraction of space in a
way dictated by the Friedmann equation \cite{Friedmann}
\begin{equation} \label{Friedmann}
  \left(\frac{\dot{a}}{a}\right)^2=\frac{8\pi}{3}G\rho(a)
\end{equation}
which is the reduction of Einstein's equations under the assumption of
isotropy. In this equation, $G$ is the gravitational constant and
$\rho(a)$ the energy density of whatever matter we have in the
universe. Once the matter content is chosen and $\rho(a)$ is known,
one can solve the Friedmann equation in order to obtain $a(t)$.

As an example we consider the case of radiation which can be
described phenomenologically by the energy density $\rho(a)\propto
a^{-4}$. This is only a phenomenological description since it ignores
the fundamental formulation of electrodynamics of the Maxwell field.
Instead of using the Maxwell Hamiltonian in order to define the energy
density, which would complicate the situation by introducing the
electromagnetic fields with new field equations coupled to the
Friedmann equation, one uses the fact that on large scales the energy
density of radiation is diluted by the expansion and in addition
red-shifted. This leads to a behavior proportional $a^{-3}$ from
dilution times $a^{-1}$ from redshift. In this example we then
solve the Friedmann equation $\dot{a}\propto a^{-1}$
by $a(t)\propto\sqrt{t-t_0}$ with a constant of integration $t_0$.
This demonstrates the occurrence of singularities: For any solution
there is a time $t=t_0$ where the size of space vanishes and the
energy density $\rho(a(t_0))$ diverges. At this point not only the
matter system becomes unphysical, but also the gravitational evolution
breaks down: When the right hand side of (\ref{Friedmann}) diverges at
some time $t_0$, we cannot follow the evolution further by setting up
an initial value problem there and integrating the equation. We can
thus only learn that there is a singularity in the classical theory,
but do not obtain any information as to what is happening there and
beyond. These are the two related but not identical features of a
singularity: energy densities diverge and the evolution breaks down.

One could think that the problem comes from too strong idealizations
such as symmetry assumptions or the phenomenological description of
matter. That this is not the case follows from the singularity
theorems which do not depend on these assumptions. One can also
illustrate the singularity problem with a field theoretic rather than
phenomenological description of matter. For simplicity we now assume
that matter is provided by a scalar $\phi$ whose energy density then
follows from the Hamiltonian
\begin{equation} \label{Hscalar}
  \rho(a)=a^{-3}H(a)= a^{-3}({\textstyle\frac{1}{2}}a^{-3}p_{\phi}^2+
a^3V(\phi))
\end{equation}
with the scalar momentum $p_{\phi}$ and potential $V(\phi)$. At small
scale factors $a$, there still is a diverging factor $a^{-3}$ in the
kinetic term which we recognized as being responsible for the
singularity before. Since this term dominates over the non-diverging
potential term, we still cannot escape the singularity by using this
more fundamental description of matter. This is true unless we manage
to arrange the evolution of the scalar in such a way that
$p_{\phi}\to0$ when $a\to0$ in just the right way for the kinetic term
not to diverge. This is difficult to arrange in general, but is
exactly what is attempted in slow-roll inflation (though with a
different motivation, and not necessarily all the way up to the
classical singularity).

For the evolution of $p_{\phi}$ we need the scalar equation of motion,
which can be derived from the Hamiltonian $H$ in (\ref{Hscalar}) via
$\dot{\phi}=\{\phi,H\}$ and $\dot{p}_{\phi}=\{p_{\phi},H\}$. This
results in the isotropic Klein--Gordon equation in a time-dependent
background determined by $a(t)$,
\begin{equation} \label{KG}
 \ddot{\phi}+3\dot{a}a^{-1}\dot{\phi}+V'(\phi)=0\,.
\end{equation}
In an expanding space with positive $\dot{a}$ the second term implies
friction such that, if we assume the potential $V'(\phi)$ to be flat
enough, $\phi$ will change only slowly (slow-roll). Thus, $\dot{\phi}$
and $p_{\phi}=a^3\dot{\phi}$ are small and at least for some time we
can ignore the kinetic term in the energy density. Moreover, since
$\phi$ changes only slowly we can regard the potential $V(\phi)$ as a
constant $\Lambda$ which again allows us to solve the Friedmann
equation with $\rho(a)=\Lambda$. The solution $a\propto
\exp(\sqrt{8\pi G\Lambda/3}\,t)$ is inflationary since $\ddot{a}>0$ and
non-singular: $a$ becomes zero only in the limit $t\to-\infty$.

Thus, we now have a mechanism to drive a phase of accelerated
expansion important for observations of structure. However, this
expansion must be long enough, which means that the phase of slowly
rolling $\phi$ must be long. This can be achieved only if the
potential is very flat and $\phi$ starts sufficiently far away from
its potential minimum. Flatness means that the ratio of $V(\phi_{\rm
  initial})$ and $\phi_{\rm initial}$ must be of the order $10^{-10}$,
while $\phi_{\rm initial}$ must be huge, of the order of the Planck
mass \cite{ChaoticInfl}. These assumptions are necessary for agreement
with observations, but are in need of more fundamental explanations.

Moreover, inflation alone does not solve the singularity
problem \cite{InflSing}. The non-singular solution we just
obtained was derived under the approximation that the kinetic term can
be ignored when $\dot{\phi}$ is small. This is true in a certain range
of $a$, depending on how small $\dot{\phi}$ really is, but never very
close to $a=0$. Eventually, even with slow-roll conditions, the
diverging $a^{-3}$ will dominate and lead to a singularity.

\section{Quantum Gravity}

For decades, quantum gravity has been expected to complete the
picture which is related to well-known properties of quantum mechanics
in the presence of a non-zero $\hbar$.

\subsection{Indications}

First, in analogy to the singularity problem in gravity, where
everything falls into a singularity in finite time, there is the
instability problem of a classical hydrogen atom, where the electron
would fall into the nucleus after a brief time. From quantum mechanics
we know how the instability problem is solved: There is a finite
ground state energy $E_0=-\frac{1}{2} me^4/\hbar^2$, implying that the
electron cannot radiate away all its energy and not fall further once
it reaches the ground state. From the expression for $E_0$ one can see
that quantum theory with its non-zero $\hbar$ is essential for this to
happen: When $\hbar\to0$ in a classical limit, $E_0\to-\infty$ which
brings us back to the classical instability. One expects a similar
role to be played by the Planck length $\lP=\sqrt{8\pi G\hbar/c^3}\approx
10^{-35}{\rm m}$ which is tiny but non-zero in quantum theory. If,
just for dimensional reasons, densities are bounded by $\lP^{-3}$,
this would be finite in quantum gravity but diverge in the classical
limit.

Secondly, a classical treatment of black body radiation suggests the
Rayleigh--Jeans law according to which the spectral density behaves as
$\rho(\lambda)\propto\lambda^{-4}$ as a function of the wave
length. This is unacceptable since the divergence at small wave
lengths leads to an infinite total energy. Here, quantum mechanics
solves the problem by cutting off the divergence with Planck's formula
which has a maximum at a wave length $\lambda_{\rm max}\sim h/kT$ and
approaches zero at smaller scales. Again, in the classical limit
$\lambda_{\max}$ becomes zero and the expression diverges.

In cosmology the situation is similar for matter in the whole universe
rather than a cavity. Energy densities as a function of the scale
factor behave as, e.g., $a^{-3}$ if matter is just diluted or $a^{-4}$
if there is an additional redshift factor. In all cases, the energy
density diverges at small scales, comparable to the Rayleigh--Jeans
law. Inflation already provides an indication that the behavior must
be different at small scales. Indeed, inflation can only be achieved
with negative pressure, while all matter whose energy falls off as
$a^{-k}$ with non-negative $k$ has positive pressure. This can easily
be seen from the thermodynamical definition of pressure as the
negative change of energy with volume. Negative pressure then requires
the energy to increase with the scale factor at least at small scales
where inflation is required (e.g., an energy $\Lambda a^3$ for
exponential inflation). This could be reconciled with standard forms
of matter if there is an analog to Planck's formula, which
interpolates between decreasing behavior at large scales and a
behavior increasing from zero at small scales, with a maximum in
between.

\subsection{Early Quantum Cosmology}

Since the isotropic reduction of general relativity leads to a system
with finitely many degrees of freedom, one can in a first attempt try
quantum mechanics to quantize it. Starting with the Friedmann equation
(\ref{Friedmann}) and replacing $\dot{a}$ by its momentum
$p_a=3a\dot{a}/8\pi G$ gives a Hamiltonian which is quadratic in
the momentum and can be quantized easily to an operator acting on a
wave function depending on the gravitational variable $a$ and possibly
matter fields $\phi$. The usual Schr\"odinger representation yields
the Wheeler--DeWitt equation \cite{DeWitt,QCReview}
\begin{equation}\label{WdW}
 \frac{3}{2}\left(-\frac{1}{9}\lP^4
  a^{-1}\frac{\partial}{\partial a}a^{-1}
  \frac{\partial}{\partial a}\right) a
\psi(a,\phi)= 8\pi G\hat{H}_{\phi}(a) \psi(a,\phi)
\end{equation}
with the matter Hamiltonian $\hat{H}_{\phi}(a)$. This system is
different from usual quantum mechanics in that there are factor
ordering ambiguities in the kinetic term, and that there is no
derivative with respect to coordinate time $t$. The latter fact is a
consequence of general covariance: the Hamiltonian is a constraint
equation restricting allowed states $\psi(a,\phi)$, rather than a
Hamiltonian generating evolution in coordinate time.  Nevertheless,
one can interpret equation (\ref{WdW}) as an evolution equation in the
scale factor $a$, which is then called internal time.  The left hand
side thus becomes a second order time derivative, and it means that
the evolution of matter is measured relationally with respect to the
expansion or contraction of the universe, rather than absolutely in
coordinate time.

Straightforward quantization thus gives us a quantum evolution
equation, and we can now check what this implies for the singularity.
If we look at the equation for $a=0$, we notice first that the matter
Hamiltonian still leads to diverging energy densities. If we quantize
(\ref{Hscalar}), we replace $p_{\phi}$ by a derivative, but the
singular dependence on $a$ does not change; $a^{-3}$ would simply
become a multiplication operator acting on the wave function. Moreover,
$a=0$ remains a singular point of the quantum evolution equation in
internal time. There is nothing from the theory which tells us what
physically happens at the singularity or beyond (baring intuitive
pictures which have been developed from this
perspective \cite{tunneling,nobound}).

So one has to ask what went wrong with our expectations that
quantizing gravity should help. The answer is that quantum theory
itself did not necessarily fail, but only our simple implementation.
Indeed, what we used was just quantum mechanics, while quantum gravity
has many consistency conditions to be fulfilled which makes
constructing it so complicated. At the time when this formalism was
first applied there was in fact no corresponding full quantum theory
of gravity which could have guided developments. In such a simple case
as isotropic cosmology, most of these consistency conditions
trivialize and one can easily overlook important issues. There are
many choices in quantizing an unknown system, and tacitly making one
choice can easily lead in a wrong direction.

Fortunately, the situation has changed with the development of
strong candidates for quantum gravity. This then allows us to
reconsider the singularity and other problems from the point of view
of the full theory, making sure that also in a simpler cosmological
context only those steps are undertaken that have an analog in the full
theory.

\subsection{Loop quantum gravity}

Singularities are physically extreme and require special properties of
any theory aimed at tackling them. First, there are always strong
fields (classically diverging) which requires a non-perturbative
treatment. Moreover, classically we expect space to degenerate at the
singularity, for instance a single point in an isotropic model. This
means that we cannot take the presence of a classical geometry to
measure distances for granted, which is technically expressed as
background independence. A non-perturbative and background independent
quantization of gravity is available in the form of loop quantum
gravity \cite{Rov:Loops,ThomasRev,ALRev}, which by now is understood
well enough in order to be applicable in physically interesting
situations.

Here, we only mention salient features of the theory which will turn
out to be important for cosmology. The first one is the kind of basic
variables used, which are the Ashtekar
connection \cite{AshVar,AshVarReell} describing the curvature of space
and a densitized triad describing the metric by a collection of three
orthonormal vectors in each point. These variables are important since
they allow a background independent representation of the theory,
where the connection $A_a^i$ is integrated to holonomies
\begin{equation} \label{Hol}
 h_e(A)={\cal P}\exp\int_e A_a^i\tau_i \dot{e}^a\md t
\end{equation}
along curves $e$ in space and the densitized triad $E^a_i$ to fluxes
\begin{equation} \label{Flux}
 F_S(E)=\int_S E^a_i\tau^in_a\md^2y
\end{equation}
along surfaces $S$. (In these expressions, $\dot{e}^a$ denotes the
tangent vector to a curve and $n_a$ the conormal to a surface, both of
which are defined without reference to a background metric. Moreover,
$\tau_j=-\frac{1}{2}i\sigma_j$ in terms of Pauli matrices). While
usual quantum field theory techniques rest on the presence of a
background metric, for instance in order to decompose a field in its
Fourier modes and define a vacuum state and particles, this is no
longer available in quantum gravity where the metric itself must be
turned into an operator. On the other hand, some integration is
necessary since the fields themselves are distributional in quantum
field theory and do not allow a well-defined representation. This
``smearing'' with respect to a background metric has to be replaced by
some other integration sufficient for resulting in honest
operators \cite{LoopRep,ALMMT}. This is achieved by the integrations in
(\ref{Hol}) and (\ref{Flux}), which similarly lead to a well-defined
quantum representation. Usual Fock spaces in perturbative quantum
field theory are thereby replaced by the loop representation, where an
orthonormal basis is given by spin network states \cite{RS:Spinnet}.

This shows that choosing basic variables for a theory to quantize has
implications for the resulting representation. Connections and
densitized triads can naturally be smeared along curves and surfaces
without using a background metric and then represented on a Hilbert
space. Requiring diffeomorphism invariance, which means that a
background independent theory must not change under deformations of
space (which can be interpreted as changes of coordinates), even
selects a unique
representation \cite{FluxAlg,Meas,HolFluxRep,SuperSel,WeylRep}. These are
basic properties of loop quantum gravity, recognized as important
requirements for a background independent quantization. Already here
we can see differences to the Wheeler--DeWitt quantization, where the
metric is used as a basic variable and then quantized as in quantum
mechanics. This is possible in the model but not in a full theory, and
in fact we will see later that a loop quantization will give a
representation inequivalent to the Wheeler--DeWitt quantization.

The basic properties of the representation have further consequences.
Holonomies and fluxes act as well-defined operators, and fluxes have
discrete spectra. Since spatial geometry is determined by the
densitized triad, spatial geometry is discrete, too, with discrete
spectra for, e.g., the area and volume
operator \cite{AreaVol,Area,Vol2}. The geometry of space-time is more
complicated to understand since this is a dynamical situation which
requires solving the Hamiltonian constraint. This is the analog of the
Wheeler--DeWitt equation in the full theory and is the quantization of
Einstein's dynamical equations. There are candidates for such
operators \cite{QSDI}, well-defined even in the presence of
matter \cite{QSDV} which in usual quantum field theory would contribute
divergent matter Hamiltonians. Not surprisingly, the full situation is
hard to analyze, which is already the case classically, without assuming
simplifications from symmetries. We will thus return to symmetric, in
particular isotropic models, but with the new perspective provided by
the full theory of loop quantum gravity.

\section{Quantum cosmology}

Symmetries can be introduced in loop quantum gravity at the level of
states and basic operators \cite{PhD,SymmRed,SphSymm}, such that it is not
necessary to reduce the classical theory first and then quantize as in
the Wheeler--DeWitt quantization. Instead, one can view the procedure
as quantizing first and then introducing symmetries which ensures that
consistency conditions of quantum gravity are observed in the first
step before one considers treatable situations. In particular, the
quantum representation derives from symmetric states and basic
operators, while the Hamiltonian constraint can be obtained with
constructions analogous to those in the full theory. This allows us to
reconsider the singularity problem, now with methods from full quantum
gravity. In fact, symmetric models present a class of systems which
can often be treated explicitly while still being representative for
general phenomena. For instance, the prime examples of singular
situations in gravity, and some of the most widely studied physical
applications, are already obtained in isotropic or spherically
symmetric systems, which allow access to cosmology and black holes.

\subsection{Representation}

Before discussing the quantum level we reformulate isotropic cosmology
in connection and triad variables instead of $a$. The role of the
scale factor is now played by the densitized triad component $p$ with
$|p|=a^2$ whose canonical momentum is the isotropic connection
component $c=-\frac{1}{2}\dot{a}$ with $\{c,p\}=8\pi G/3$. The main
difference to metric variables is the fact that $p$, unlike $a$, can
take both signs with $\sgn p$ being the orientation of space. This is
a consequence of having to use triad variables which not only know
about the size of space but also its orientation (depending on whether
the set of orthonormal vectors is left or right handed).

States in the full theory are usually written in the connection
representation as functions of holonomies. Following the reduction
procedure for an isotropic symmetry group leads to orthonormal states which are
functions of the isotropic connection component $c$ and given by \cite{Bohr}
\begin{equation} \label{basis}
 \langle c|\mu\rangle= e^{i\mu c/2} \qquad \mu\in{\mathbb R}\,.
\end{equation}
On these states the basic variables $p$ and $c$ are represented by
\begin{eqnarray}
\hat{p}|\mu\rangle &=& {\textstyle\frac{1}{6}}\lP^2\mu|\mu\rangle\\
\widehat{e^{i\mu'c/2}}|\mu\rangle &=& |\mu+\mu'\rangle 
 \end{eqnarray}
with the properties:
\begin{enumerate}
 \item $[\widehat{e^{i\mu'
       c/2}},\hat{p}]=-\frac{1}{6}\lP^2\mu'\widehat{e^{-i\mu' c/2}}=
   i\hbar(\{e^{i\mu' c/2},p\})^{\wedge}$,
 \item $\hat{p}$ has a discrete spectrum and
 \item only exponentials $e^{i\mu'c/2}$ of $c$ are represented, not
   $c$ directly.
\end{enumerate}
These statements deserve further explanation: First, the classical
Poisson relations between the basic variables are indeed represented
correctly, turning the Poisson brackets into commutators divided by
$i\hbar$.  On this representation, the set of eigenvalues of $\hat{p}$
is the full real line since $\mu$ can take arbitrary real values.
Nevertheless, the spectrum of $\hat{p}$ is discrete in the technical
sense that eigenstates of $\hat{p}$ are normalizable. This is indeed
the case in this non-separable Hilbert space where (\ref{basis})
defines an orthonormal basis. The last property follows since the
exponentials are not continuous in the label $\mu'$, for otherwise one
could simply take the derivative with respect to $\mu'$ at $\mu'=0$
and obtain an operator for $c$. The discontinuity can be seen, e.g., from
\[
 \langle\mu|\widehat{e^{i\mu'c/2}}|\mu\rangle= \delta_{0,\mu'}
\]
which is not continuous.

These properties are quite unfamiliar from quantum mechanics, and
indeed the representation is inequivalent to the Schr\"odinger
representation (the discontinuity of the $c$-exponential evading the
Stone--von Neumann theorem which usually implies uniqueness of the
representation). In fact, the loop representation is inequivalent to
the Wheeler-DeWitt quantization which just assumed a Schr\"odinger
like quantization. In view of the fact that the phase space of our
system is spanned by $c$ and $p$ with $\{c,p\}\propto 1$ just as in
classical mechanics, the question arises how such a difference in the
quantum formulation arises.

As a mathematical problem the basic step of quantization occurs as
follows: given the classical Poisson algebra of observables $Q$ and
$P$ with $\{Q,P\}=1$, how can we define a representation of the
observables on a Hilbert space such that the Poisson relations become
commutator relations and complex conjugation, meaning that $Q$ and $P$
are real, becomes adjointness? The problem is mathematically much
better defined if one uses the bounded expressions $e^{isQ}$ and
$e^{it\hbar^{-1}P}$ instead of the unbounded $Q$ and $P$, which still
allows us to distinguish any two points in the whole phase space. The
basic objects $e^{isQ}$ and $e^{it\hbar^{-1}P}$ upon quantization will
then not commute but fulfill the commutation relation (Weyl algebra)
\begin{equation}\label{QP}
 e^{isQ}e^{it\hbar^{-1}P} = e^{ist} e^{it\hbar^{-1}P} e^{isQ}
\end{equation}
as unitary operators on a Hilbert space.

In the Schr\"odinger representation this is done by using a Hilbert
space $L^2({\mathbb R},\md q)$ of square integrable functions
$\psi(q)$ with $\int_{\mathbb R}\md q|\psi(q)|^2$ finite. The
representation of basic operators is
\begin{eqnarray*}
 e^{isQ}\psi(q) &=& e^{isq}\psi(q)\\
 e^{it\hbar^{-1}P}\psi(q) &=& \psi(q+t)
\end{eqnarray*}
which indeed are unitary and fulfill the required commutation
relation. Moreover, the operator families as functions of $s$ and $t$
are continuous and we can take the derivatives in $s=0$ and $t=0$,
respectively:
\begin{eqnarray*}
 -i\left.\frac{\md}{\md s}\right|_{s=0} e^{isQ} &=& q\\
 -i\hbar\left.\frac{\md}{\md t}\right|_{t=0} e^{it\hbar^{-1}P} &=&
 \hat{p}=-i\hbar\frac{\md}{\md q}\,.
\end{eqnarray*}
This is the familiar representation of quantum mechanics which,
according to the Stone--von Neumann theorem is unique under the
condition that $e^{isQ}$ and $e^{it\hbar^{-1}P}$ are indeed continuous
in both $s$ and $t$.

The latter condition is commonly taken for granted in quantum
mechanics, but in general there is no underlying physical or
mathematical reason. It is easy to define representations violating
continuity in $s$ or $t$, for instance if we use a Hilbert space
$\ell^2({\mathbb R})$ where states are again maps $\psi_q$ from the
real line to complex numbers but with norm $\sum_q|\psi_q|^2$ which
implies that normalizable $\psi_q$ can be non-zero for at most
countably many $q$. We obtain a representation with basic operators
\begin{eqnarray*}
 e^{isQ}\psi_q &=& e^{isq}\psi_q\\
 e^{it\hbar^{-1}P}\psi_q &=& \psi_{q+t}
\end{eqnarray*}
which is of the same form as before. However, due to the different
Hilbert space the second operator $e^{it\hbar^{-1}P}$ is no longer
continuous in $t$ which can be checked as in the case of $e^{i\mu
  c/2}$. In fact, the representation for $Q$ and $P$ is isomorphic to
that of $p$ and $c$ used before, where a general state
$|\psi\rangle=\sum_{\mu}\psi_{\mu}|\mu\rangle$ has coefficients
$\psi_{\mu}$ in $\ell^2({\mathbb R})$. 

This explains mathematically why different, inequivalent
representations are possible, but what are the physical reasons for
using different representations in quantum mechanics and quantum
cosmology? In quantum mechanics it turns out that the choice of
representation is not that important and is mostly being done for
reasons of familiarity with the standard choice. Physical differences
between inequivalent representations can only occur at very high
energies \cite{PolymerParticle} which are not probed by available
experiments and do not affect characteristic quantum effects related
to the ground state or excited states. Thus, quantum mechanics as we
know it can well be formulated in an inequivalent representation, and
also in quantum field theory this can be done and even be
useful \cite{QEDBohr}.

In quantum cosmology we have a different situation where it is the high
energies which are essential. We do not have direct observations of
this regime, but from conceptual considerations such as the
singularity issue we have learned which problems we have to face. The
classical singularity leads to the highest energies one can imagine,
and it is here where the question of which representation to choose
becomes essential. As shown by the failure of the Wheeler--DeWitt
quantization in trying to remove the singularity, the Schr\"odinger
representation is inappropriate for quantum cosmology. The
representation underlying loop quantum cosmology, on the other hand,
implies very different properties which become important at high
energies and can shed new light on the singularity problem.

Moreover, by design of the symmetric models as derived from the full
theory, we have the same basic properties of a loop representation in
cosmological models and the full situation where they were recognized
as being important for a background independent quantization: discrete
fluxes $\hat{F}_S(E)$ or $\hat{p}$ and a representation only of
holonomies $h_e(A)$ or $e^{i\mu c/2}$ but not of connection components
$A_a^i$ or $c$. These basic properties have far-reaching consequences
as discussed in what follows \cite{LoopCosRev}:
\begin{center}
\begin{picture}(160,80)(0,0)
\put(80,70){\makebox(0,0){discrete triad \hspace{2cm} only holonomies}}
\put(30,60){\vector(0,-1){10}} \put(130,60){\vector(0,-1){10}}
\put(80,45){\makebox(0,0){finite inverse volume \hspace{2cm} discrete evolution}}
\put(40,40){\vector(1,-1){10}} \put(125,40){\vector(-1,-1){10}}
\put(80,25){\makebox(0,0){ non-singular}}
\put(30,35){\vector(0,-1){20}} \put(130,35){\vector(0,-1){20}}
\put(80,5){\makebox(0,0){\hspace{-1cm}non-perturbative modifications \hspace{1cm}
higher order terms}}
\end{picture}
\end{center}
By this reliable quantization of representative and physically
interesting models with a known relation to full quantum gravity we
are finally able to resolve long-standing issues such as the
singularity problem.

\subsection{Quantum evolution}

We will first look at the quantum evolution equation which we obtain as
the quantized Friedmann equation. This is modeled on the Hamiltonian
constraint of the full theory such that we can also draw some
conclusions for the viability of the full constraint.

\subsubsection{Difference equation}

The constraint equation will be imposed on states of the form
$|\psi\rangle=\sum_{\mu}\psi_{\mu}|\mu\rangle$ with summation over
countably many values of $\mu$. Since the states $|\mu\rangle$ are
eigenstates of the triad operator, the coefficients $\psi_{\mu}$ which
can also depend on matter fields such as a scalar $\phi$ represent the
state in the triad representation, analogous to $\psi(a,\phi)$ before.
For the constraint operator we again need operators for the conjugate
of $p$, related to $\dot{a}$ in the Friedmann equation. Since this is
now the exponential of $c$, which on basis states acts by shifting the
label, it translates to a finite shift in the labels of coefficients
$\psi_{\mu}(\phi)$. Plugging together all ingredients for a
quantization of (\ref{Friedmann}) along the lines of the constraint in
the full theory leads to the difference equation \cite{IsoCosmo,Bohr}
\begin{eqnarray} \label{DiffEq}
&&    (V_{\mu+5}-V_{\mu+3})\psi_{\mu+4}(\phi)- 2
(V_{\mu+1}-V_{\mu-1})\psi_{\mu}(\phi)\\\nonumber
&&+    (V_{\mu-3}-V_{\mu-5})\psi_{\mu-4}(\phi)
  = -{\textstyle\frac{4}{3}}\pi
G\ell_{\rm P}^2\hat{H}_{\rm matter}(\mu)\psi_{\mu}(\phi)
\end{eqnarray}
with volume eigenvalues $V_{\mu}=(\ell_{\rm P}^2|\mu|/6)^{3/2}$
obtained from the volume operator $\hat{V}=|\hat{p}|^{3/2}$, and the
matter Hamiltonian $\hat{H}_{\rm matter}(\mu)$.

We again have a constraint equation which does not generate evolution
in coordinate time but can be seen as evolution in internal time.
Instead of the continuous variable $a$ we now have the label $\mu$
which only jumps in discrete steps. As for the singularity issue,
there is a further difference to the Wheeler--DeWitt equation since
now the classical singularity is located at $p=0$ which is in the
interior rather than at the boundary of the configuration space.
Nevertheless, the classical evolution in the variable $p$ breaks down
at $p=0$ and there is still a singularity.  In quantum theory,
however, the situation is very different: while the Wheeler--DeWitt
equation does not solve the singularity problem, the difference
equation (\ref{DiffEq}) uniquely evolves a wave function from some
initial values at positive $\mu$, say, to negative $\mu$. Thus, the
evolution does not break down at the classical singularity and can
rather be continued beyond it. Quantum gravity is thus a theory which
is more complete than classical general relativity and is free of
limitations set by classical singularities.

An intuitive picture of what replaces the classical singularity can be
obtained from considering evolution in $\mu$ as before. For negative
$\mu$, the volume $V_{\mu}$ decreases with increasing $\mu$ while
$V_{\mu}$ increases for positive $\mu$. This leads to the picture of a
collapsing universe before it reaches the classical big bang
singularity and re-expands. While at large scales the classical
description is good \cite{SemiClass}, when the universe is small close
to the classical singularity it starts to break down and has to be
replaced by discrete quantum geometry. The resulting quantum
evolution does not break down, in contrast to the classical space-time
picture which dissolves. Using the fact that the sign of $\mu$, which
defines the orientation of space, changes during the transition
through the classical singularity one can conclude that the universe
turns its inside out during the process. This can have consequences
for realistic matter Hamiltonians which violate parity symmetry.

\subsubsection{Meaning of the wave function}
\label{sec:wavefct}

An important issue in quantum gravity which is still outstanding even
in isotropic models is the interpretation of the wave function and its
relation to the problem of time. In the usual interpretation of
quantum mechanics the wave function determines probabilities for
measurements made by an observer outside the quantum system. Quantum
gravity and cosmology, however, are thought of as theories for the
quantum behavior of a whole universe such that, by definition, there
cannot be an observer outside the quantum system. Accordingly, the
question of how to interpret the wave function in quantum cosmology is
more complicated. One can avoid the separation into a classical and
quantum part of the problem in quantum mechanics by the theory of
decoherence which can explain how a world perceived as classical
emerges from the fundamental quantum description \cite{Decoherence}.
The degree of ``classicality'' is related to the number of degrees of
freedom which do not contribute significantly to the evolution but
interact with the system nonetheless. Averaging over those degrees of
freedom, provided there are enough of them, then leads to a classical
picture. This demonstrates why macroscopic bodies under usual
circumstances are perceived as classical while in the microscopic
world, where a small number of degrees of freedom is sufficient to
capture crucial properties of a system, quantum mechanics prevails.
This idea has been adapted to cosmology, where a large universe comes
with many degrees of freedom such as small inhomogeneities which are
not of much relevance for the overall evolution. This is different,
however, in a small universe where quantum behavior becomes dominant.

Thus, one can avoid the presence of an observer outside the quantum
system. The quantum system is described by its wave function, and in
some circumstances one can approximate the evolution by a quantum part
being looked at by classical observers within the same
system. Properties are then encoded in a relational way: the wave
function of the whole system contains information about everything
including possible observers. Now, the question has shifted from a
conceptual one --- how to describe the system if no outside observers
can be available --- to a technical one. One needs to understand how
information can be extracted from the wave function and used to
develop schemes for intuitive pictures or potentially observable
effects. This is particularly pressing in the very early universe
where everything including what we usually know as space and time are
quantum and no familiar background to lean on is left.

One lesson is that evolution should be thought of as relational by
determining probabilities for one degree of freedom under the
condition that another degree of freedom has a certain value. If the
reference degree of freedom (such as the direction of the hand of a
clock) plays a distinguished role for determining the evolution of
others, it is called internal time: it is not an absolute time outside
the quantum system as in quantum mechanics, and not a coordinate time
as in general relativity which could be changed by coordinate
transformations. Rather, it is one of the physical degrees of freedom
whose evolution is determined by the dynamical laws and which shows
how other degrees of freedom change by interacting with them. From
this picture it is clear that no external observer is necessary to
read off the clock or other measurement devices, such that it is
ideally suited to cosmology. What is also clear is that now internal
time depends on what we choose it to be, and different questions
require different choices. For a lab experiment the hand of a clock
would be a good internal time and, when the clock is sufficiently
isolated from the physical fields used in the experiment and other
outside influence, will not be different from an absolute time except
that it is mathematically more complicated to describe. The same
clock, on the other hand, will not be good for describing the universe
when we imagine to approach a classical singularity. It will simply
not withstand the extreme physical conditions, dissolve, and its parts
will behave in a complicated irregular manner ill-suited for the
description of evolution. Instead, one has to use more global objects
which depend on what is going on in the whole universe.

Close to a classical singularity, where one expects monotonic
expansion or contraction, the spatial volume of the universe is just
the right quantity as internal time. A wave function then determines
relationally how matter fields or other gravitational degrees of
freedom change with respect to the expansion or contraction of the
universe. In our case, this is encoded in the wave function
$\psi_{\mu}(\phi)$ depending on internal time $\mu$ (which through the
volume defines the size of the universe but also spatial orientation)
and matter fields $\phi$. By showing that it is subject to a
difference equation in $\mu$ which does not stop at the classical
singularity $\mu=0$ we have seen that relational probabilities are
defined for all internal times without breaking down anywhere. This
shows the absence of singularities and allows developing intuitive
pictures, but does not make detailed predictions before relational
probabilities are indeed computed and shown how to be observable at
least in principle. 

Here, we encounter the main issue in the role of the wave function: we
have a relational scheme to understand what the wave function should
mean but the probability measure to be used, called the physical inner
product, is not known so far. We already used a Hilbert space which we
needed to define the basic operators and the quantized Hamiltonian
constraint, where wave functions $\psi_{\mu}$, which by definition are
non-zero for at most countably many values $\mu\in{\mathbb R}$, have
the inner product $\langle\psi|\psi'\rangle=
\sum_{\mu}\bar{\psi}_{\mu}\psi'_{\mu}$. This is called the kinematical
inner product which is used for setting up the quantum theory. But
unlike in quantum mechanics where the kinematical inner product is
also used as physical inner product for the probability interpretation
of the wave function, in quantum gravity the physical inner product
must be expected to be different. This occurs because the quantum
evolution equation (\ref{DiffEq}) in internal time is a constraint
equation rather than an evolution equation in an external absolute
time parameter. Solutions to this constraint in general are not
normalizable in the kinematical inner product such that a new physical
inner product on the solution space has to be found. There are
detailed schemes for a derivation, but despite some
progress \cite{Golam,IsoSpinFoam} they are difficult to apply even in
isotropic cosmological models and research is still ongoing. An
alternative route to extract physical statements will be discussed in
Sec.~\ref{sec:pheno} together with the main results.

A related issue, which is also of relevance for the classical limit of
the theory is that of oscillations on small scales of the wave
function. Being subject to a difference equation means that
$\psi_{\mu}$ is not necessarily smooth but can change rapidly when
$\mu$ changes by a small amount even when the volume is large. In such
a regime one expects classical behavior, but small scale oscillations
imply that the wave function is sensitive to the Planck scale.  There
are also other issues related to the fact that now a difference rather
than differential equation provides the fundamental
law \cite{FundamentalDisc}. Before the physical inner product is known
one cannot say if these oscillations would imply any effect observable
today, but one can still study the mathematical problem of if and when
solutions with suppressed oscillations exist. This is easy to answer
in the affirmative for isotropic models subject to (\ref{DiffEq})
where in some cases one even obtains a unique wave
function \cite{DynIn,Essay}. However, already in other homogeneous but
anisotropic models the issue is much more complicated to
analyze \cite{GenFunc,GenFuncBI}.

In a more general situation than homogeneous cosmology there is an
additional complication even if the physical inner product would be
known. In general, it is very difficult to find an internal time to
capture the evolution of a complicated quantum system, which is called
the problem of time in general relativity. In cosmology, the volume is
a good internal time to understand the singularity, but it would not
be good for the whole history if the universe experiences a recollapse
where the volume would not be monotonic. This is even more complicated
in inhomogeneous situations such as the collapse of matter into a
black hole. Since we used internal time $\mu$ to show how quantum
geometry evolves through the classical singularity, it seems that the
singularity problem in general cannot be solved before the problem of
time is understood. Fortunately, while the availability of an internal
time simplifies the analysis, requirements on a good choice can be
relaxed for the singularity problem. An internal time provides us with
an interpretation of the constraint equation as an evolution equation,
but the singularity problem can be phrased independently of this as
the problem to extend the wave function on the space of metrics or
triads. This implies weaker requirements and also situations can be
analyzed where no internal time is known. The task then is to find
conditions which characterize a classical singularity, analogous to
$p=0$ in isotropic cosmology, and find an evolution parameter which at
least in individual parts of an inhomogeneous singularity allows to
see how the system can move through it. Inhomogeneous cases are now
under study but only partially understood so far, such that in the
next section we return to isotropic cosmology.

\subsection{Densities}

In the previous discussion we have not yet mentioned the matter
Hamiltonian on the right hand side, which diverges classically and in
the Wheeler--DeWitt quantization when we reach the singularity. If
this were the case here, the discrete quantum evolution would break
down, too. However, as we will see now the matter Hamiltonian does not
diverge, which is again a consequence of the loop representation.

\subsubsection{Quantization}

For the matter Hamiltonian we need to quantize the matter field and in
quantum gravity also coefficients such as $a^{-3}$ in the kinetic term
which now become operators, too. In the Wheeler--DeWitt quantization
where $a$ is a multiplication operator, $a^{-3}$ is unbounded and
diverges at the classical singularity. In loop quantum cosmology we
have the basic operator $\hat{p}$ which one can use to construct a
quantization of $a^{-3}$. However, a straightforward quantization
fails since, as one of the basic properties, $\hat{p}$ has a discrete
spectrum containing zero. In this case, there is no densely defined
inverse operator which one could use. This seems to indicate that the
situation is even worse: an operator for the kinetic term would not
only be unbounded but not even be well-defined. The situation is much
better, however, when one tries other quantizations which are more
indirect. For non-basic operators such as $a^{-3}$ there are usually
many ways to quantize, all starting from the same classical
expression. What we can do here, suggested by constructions in the
full theory \cite{QSDV}, is to rewrite $a^{-3}$ in a classically
equivalent way as
\[
 a^{-3}=(\pi^{-1} G^{-1}{\rm
   tr}\tau_3e^{c\tau_3}\{e^{-c\tau_3},\sqrt{V}\})^6 
\]
where we only need a readily available positive power of $\hat{p}$.
Moreover, exponentials of $c$ are basic operators, where we just used
su(2) notation $e^{c\tau_3}=\cos\frac{1}{2}c+2\tau_3\sin\frac{1}{2}c$
in order to bring the expression closer to what one would have in the
full theory, and the Poisson bracket will become a commutator in
quantum theory.

This procedure, after taking the trace, leads to a densely defined
operator for $a^{-3}$ despite the nonexistence of an inverse of
$\hat{p}$ \cite{InvScale}:
\begin{equation}
 \widehat{a^{-3}} = \left(8i\lP^{-2} (\sin{\textstyle\frac{1}{2}}c
   \sqrt{\hat{V}}
 \cos{\textstyle\frac{1}{2}}c - \cos{\textstyle\frac{1}{2}}c
 \sqrt{\hat{V}} \sin{\textstyle\frac{1}{2}}c)\right)^6\,.
\end{equation}
That this operator is indeed finite can be seen from its action on
states $|\mu\rangle$ which follows from that of the basic operators:
\begin{equation}
 \widehat{a^{-3}}|\mu\rangle = \left(4\lP^{-2}
 (\sqrt{V_{\mu+1}}-\sqrt{V_{\mu-1}}\,)\right)^6 |\mu\rangle
\end{equation}
immediately showing the eigenvalues which are all finite. In
particular, at $\mu=0$ where we would have the classical singularity
the density operator does not diverge but is zero.

This finiteness of densities finally confirms the non-singular
evolution since the matter Hamiltonian
\begin{equation} \label{HmatterQuant}
 \hat{H}_{\rm
   matter}={\textstyle\frac{1}{2}}\widehat{a^{-3}}\hat{p}_{\phi}^2+ 
  \hat{V} V(\phi)
\end{equation}
in the example of a scalar is well-defined even on the classically
singular state $|0\rangle$. The same argument applies for other
matter Hamiltonians since only the general structure of kinetic and
potential terms is used.

\subsubsection{Confirmation of indications}

The finiteness of the operator is a consequence of the loop
representation which forced us to take a detour in quantizing inverse
powers of the scale factor. A more physical understanding can be
obtained by exploiting the fact that there are quantization
ambiguities in this non-basic operator. This comes from the rewriting
procedure which is possible in many classically equivalent ways,
which all lead to different operators. Important properties such as
the finiteness and the approach to the classical limit at large volume
are robust under the ambiguities, but finer details can change. The
most important choices one can make are selecting the representation
$j$ of SU(2) holonomies before taking the trace \cite{Gaul,Ambig} and
the power $l$ of $|p|$ in the Poisson bracket \cite{ICGC}. These
values are restricted by the requirement that $j$ is a half-integer
($j=1/2$ in the above choice) and $0<l<1$ to obtain a well-defined
inverse power of $a$ ($l=3/4$ above). The resulting eigenvalues can be
computed explicitly and be approximated by the formula \cite{Ambig,ICGC}
\begin{equation} \label{effdens}
 (a^{-3})_{\rm eff}= a^{-3} p_l(a^2/a_{\rm
  max}^2)^{3/(2-2l)}
\end{equation}
where $a_{\rm max}=\sqrt{j/3}\,\ell_{\rm P}$ depends on the first
ambiguity parameter and the function
\begin{eqnarray}
 p_l(q) &=&
\frac{3}{2l}q^{1-l}\left((l+2)^{-1}
\left((q+1)^{l+2}-|q-1|^{l+2}\right)\right. \label{plq}\\
 &&\qquad- \left.(l+1)^{-1}q
\left((q+1)^{l+1}-{\rm sgn}(q-1)|q-1|^{l+1}\right)\right) \,. \nonumber
\end{eqnarray}
on the second.

\begin{figure}[th]              
\centerline{\includegraphics[width=14cm]{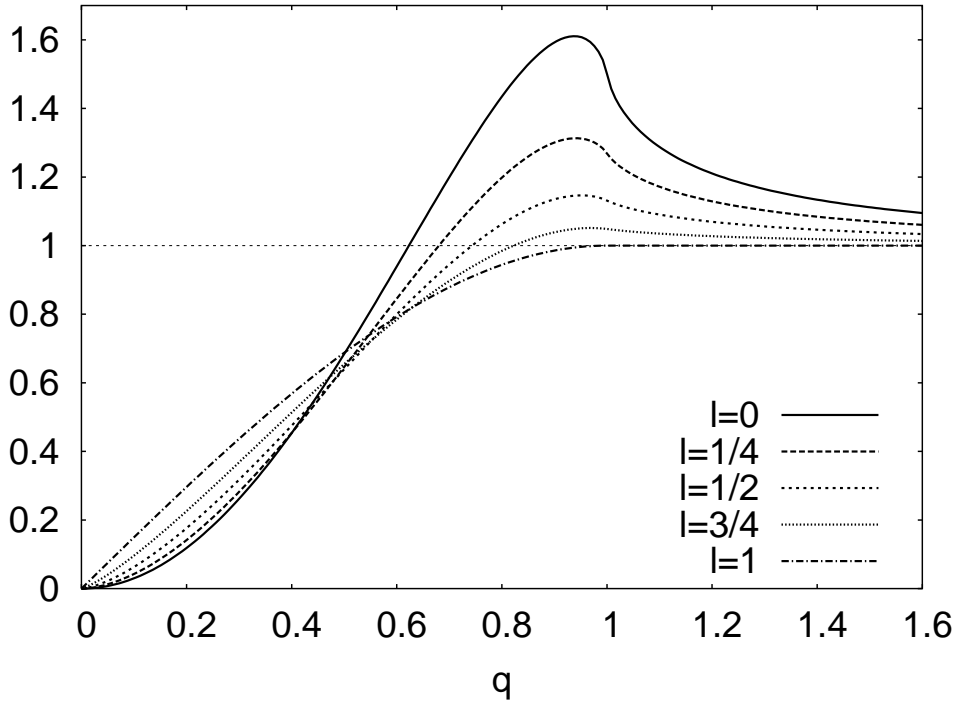}}
\vspace*{8pt}
\caption{The function $p_l(q)$ in (\ref{plq}) for some values of $l$,
  including the limiting cases $l=0$ and $l=1$.
\label{pl}}
\end{figure}

The function $p_l(q)$, shown in Fig.~\ref{pl}, approaches one for
$q\gg1$, has a maximum close to $q=1$ and drops off as $q^{2-l}$ for
$q\ll1$. This shows that $(a^{-3})_{\rm eff}$ approaches the classical
behavior $a^{-3}$ at large scales $a\gg a_{\rm max}$, has a maximum
around $a_{\rm max}$ and falls off like $(a^{-3})_{\rm eff}\sim
a^{3/(1-l)}$ for $a\ll a_{\rm max}$. The peak value can be
approximated, e.g.\ for $j=1/2$, by $(a^{-3})_{\rm eff}(a_{\rm
  max})\sim 3l^{-1}2^{-l}(1-3^{-l})^{3/(2-2l)} \lP^{-3}$ which indeed
shows that densities are bounded by inverse powers of the Planck
length such that they are finite in quantum gravity but diverge in the
classical limit.  This confirms our qualitative expectations from the
hydrogen atom, while details of the coefficients depend on the
quantization.

Similarly, densities are seen to have a peak at $a_{\rm max}$ whose
position is given by the Planck length (and an ambiguity
parameter). Above the peak we have the classical behavior of an
inverse power, while below the peak the density increases from
zero. As suggested by the behavior of radiation in a cavity whose
spectral energy density
\[
 \rho_T(\lambda)=8\pi h\lambda^{-5}(e^{h/kT\lambda}-1)^{-1} = h
 \lambda^{-5} f(\lambda/\lambda_{\rm max})
\]
can, analogously to (\ref{effdens}), be expressed as the diverging
behavior $\lambda^{-5}$ multiplied with a cut-off function
$f(y)=8\pi/((5/(5-x))^{1/y}-1)$ with $x=5+W(-5e^{-5})$ (in terms of
the Lambert function $W(x)$, the inverse function of $xe^x$) and
$\lambda_{\rm max}=h/xkT$, we obtain an interpolation between
increasing behavior at small scales and decreasing behavior at large
scales in such a way that the classical divergence is cut off.

\begin{figure}[th]              
\centerline{\includegraphics[width=15cm]{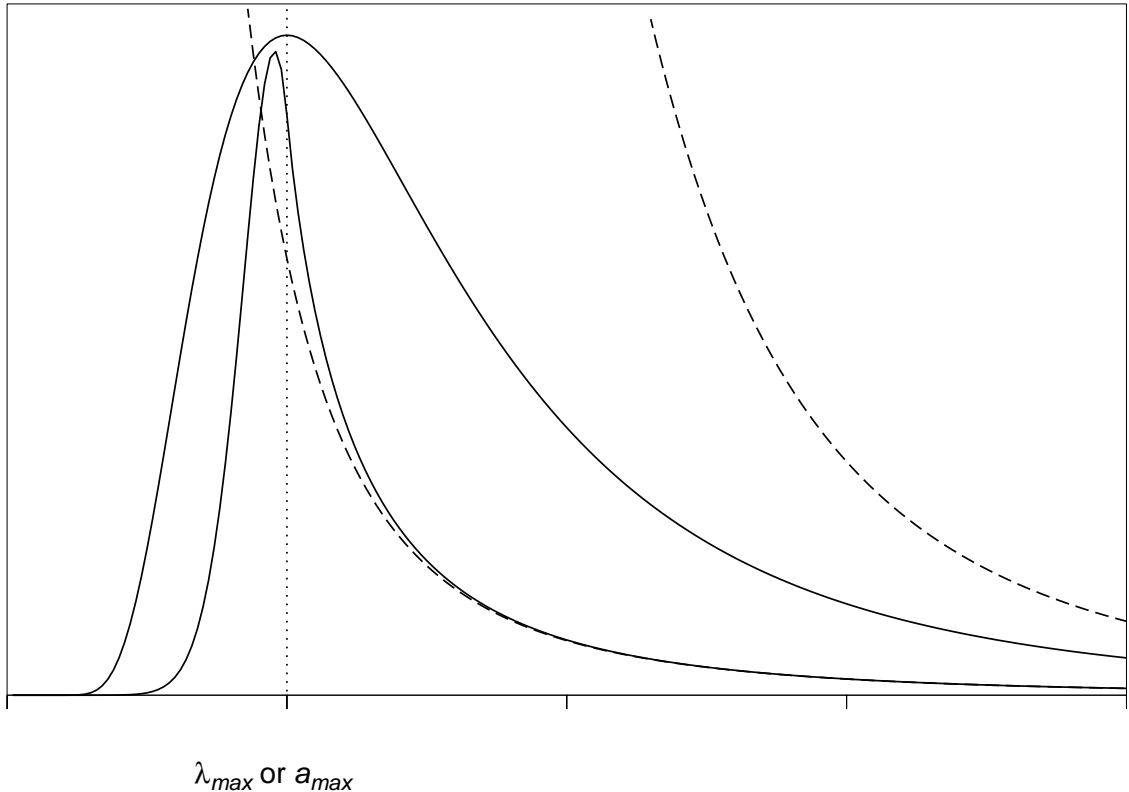}}
%\vspace*{8pt}
\caption{Comparison between the spectral energy density of black body
  radiation (wide curve) and an effective geometrical density with
  their large scale approximations (dashed).
\label{Density}}
\end{figure}

We thus have an interpolation between increasing behavior necessary
for negative pressure and inflation and the classical decreasing
behavior (Fig.~\ref{Density}). Any matter density turns to increasing
behavior at sufficiently small scales without the need to introduce an
inflaton field with tailor-made properties. In the following section
we will see the implications for cosmological evolution by studying
effective classical equations incorporating this characteristic loop
effect of modified densities at small scales.

\subsection{Phenomenology}
\label{sec:pheno}

The quantum difference equation (\ref{DiffEq}) is rather complicated
to study in particular in the presence of matter fields and, as
discussed in Sec.~\ref{sec:wavefct}, difficult to interpret in a fully
quantum regime. It is thus helpful to work with effective equations,
comparable conceptually to effective actions in field theories, which
are easier to handle and more familiar to interpret but still show
important quantum effects. This can be done
systematically \cite{Bohr,Perturb,Josh}, starting with the Hamiltonian
constraint operator, resulting in different types of correction terms
whose significance in given regimes can be estimated or studied
numerically \cite{Time}. There are perturbative corrections to the
Friedmann equation of higher order form in $\dot{a}$, or of higher
derivative, in the gravitational part on the left hand side, but also
modifications in the matter Hamiltonian since the density in its
kinetic term behaves differently at small scales. The latter
corrections are mainly non-perturbative since the full expression for
$(a^{-3})_{\rm eff}$ depends on the inverse Planck length, and their
range can be extended if the parameter $j$ is rather large. For these
reasons, those corrections are most important and we focus on them
from now on.

The effective Friedmann equation then takes the form
\begin{equation} \label{effFried}
  a\dot{a}^2={\textstyle\frac{8\pi}{3}}G
\left({\textstyle\frac{1}{2}}(a^{-3})_{\rm eff}\, p_{\phi}^2+a^3
V(\phi)\right)
\end{equation}
with $(a^{-3})_{\rm eff}$ as in (\ref{effdens}) with a choice of
ambiguity parameters. Since the matter Hamiltonian does not just act
as a source for the gravitational field on the right hand side of the
Friedmann equation, but also generates Hamiltonian equations of
motion, the modification entails further changes in the matter
equations of motion. The Klein--Gordon equation (\ref{KG}) then takes
the effective form
\begin{equation} \label{effKG}
 \ddot{\phi}=\dot{\phi}\,\dot{a}\frac{\md\log(a^{-3})_{\rm
eff}}{\md a}-a^3(a^{-3})_{\rm eff}V'(\phi)
\end{equation}
and finally there is the Raychaudhuri equation
\begin{equation} \label{effRay}
 \frac{\ddot{a}}{a}= -\frac{8\pi G}{3}\left( a^{-3}d(a)_{\rm
eff}^{-1}\dot{\phi}^2 
\left(1-{\textstyle\frac{1}{4}}a\frac{\md
\log(a^3d(a)_{\rm eff})}{\md a}\right) -V(\phi)\right)
\end{equation}
which follows from the above equation and the continuity equation of
matter.

\subsubsection{Bounces}

The resulting equations can be studied numerically or with qualitative
analytic techniques. We first note that the right hand side of
(\ref{effFried}) behaves differently at small scales since it
increases with $a$ at fixed $\phi$ and $p_{\phi}$. Viewing this
equation as analogous to a constant energy equation in classical
mechanics with kinetic term $\dot{a}^2$ and potential term ${\cal
  V}(a):=-{\textstyle\frac{8\pi}{3}}G a^{-1}
\left({\textstyle\frac{1}{2}}(a^{-3})_{\rm eff}\, p_{\phi}^2+a^3
  V(\phi)\right)$ illustrates the classically attractive nature of
gravity: The dominant part of this potential behaves like $-a^{-4}$
which is increasing. Treating the scale factor analogously to the
position of a classical particle shows that $a$ will be driven toward
smaller values, implying attraction of matter and energy in the
universe. This changes when we approach smaller scales and take into
account the quantum modification. Below the peak of the effective
density the classical potential ${\cal V}(a)$ will now decrease,
$-{\cal V}(a)$ behaving like a positive power of $a$. This implies
that the scale factor will be repelled away from $a=0$ such that there
is now a small-scale repulsive component to the gravitational force if
we allow for quantum effects. The collapse of matter can then be
prevented if repulsion is taken into account, which indeed can be
observed in some models where the effective classical equations alone
are sufficient to demonstrate singularity-free evolution.

This happens by the occurrence of bounces where $a$ turns around from
contracting to expanding behavior. Thus, $\dot{a}=0$ and $\ddot{a}>0$.
The first condition is not always realizable, as follows from the
Friedmann equation (\ref{Friedmann}). In particular, when the scalar
potential is non-negative there is no bounce, which is not changed by
the effective density. There are then two possibilities for bounces in
isotropic models, the first one if space has positive curvature rather
than being flat as assumed here \cite{BounceClosed,BounceQualitative},
the second one with a scalar potential which can become
negative \cite{Oscill,Cyclic}. Both cases allow $\dot{a}=0$
even in the classical case, but this always corresponds to a maximum
rather than minimum. This can easily be seen for the case of negative
potential from the Raychaudhuri equation (\ref{effRay}) which in the
classical case implies negative $\ddot{a}$. With the modification,
however, the additional term in the equation provides a positive
contribution which can become large enough for $\ddot{a}$ to become
positive at a value of $\dot{a}=0$ such that there is a bounce.

This provides intuitive explanations for the absence of singularities
from quantum gravity, but not a general one. The generic presence of
bounces depends on details of the model such as its matter content or
which correction terms are being used \cite{EffHam,GenericBounce}, and
even with the effective modifications there are always models which
classically remain singular. Thus, the only general argument for
absence of singularities remains the quantum one based on the
difference equation (\ref{DiffEq}), where the conclusion is model
independent.

\subsubsection{Inflation}

\begin{figure}[th]              
\centerline{\includegraphics[width=16cm]{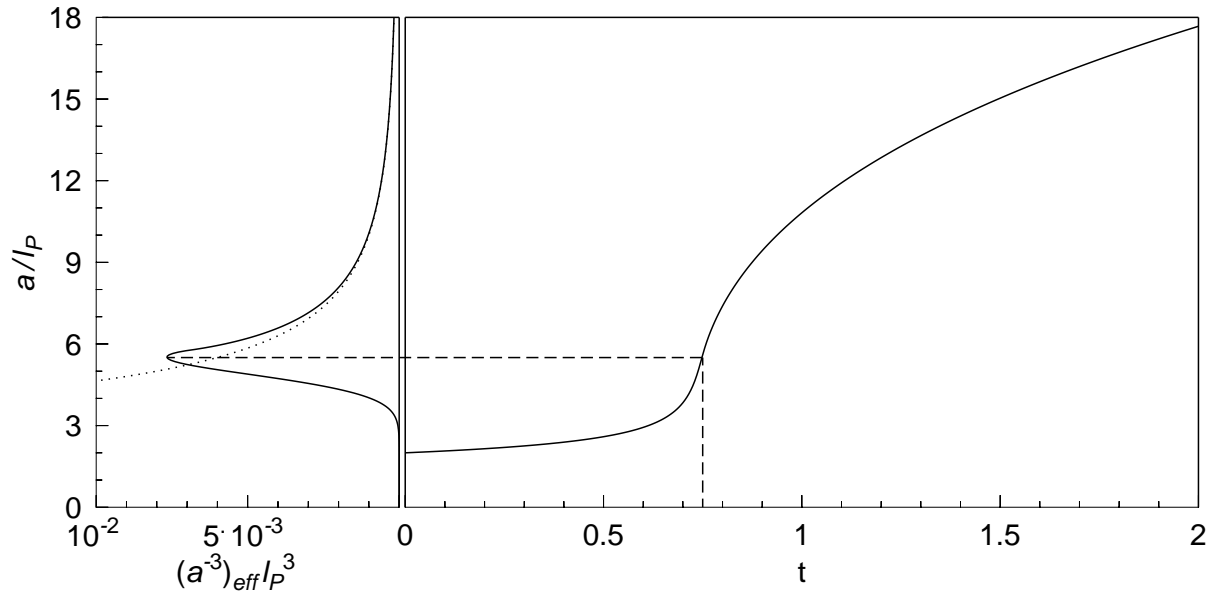}}
\vspace*{8pt}
\caption{Numerical solution to the effective Friedmann equation
  (\ref{effFried}) with a vanishing scalar potential. While the
  modification in the density on the left is active the expansion is
  accelerated, which stops automatically once the universe expands to
  a size above the peak in the effective density.
\label{InflAeff}}
\end{figure}

A repulsive contribution to the gravitational force can not only
explain the absence of singularities, but also enhances the expansion
of the universe on scales close to the classical singularity. Thus, as
seen also in Fig.~\ref{InflAeff} the universe accelerates just from
quantum effects, providing a mechanism for inflation without choosing
special matter.

Via the generation of structure, inflationary phases of the universe
can have an imprint on the observable cosmic microwave background.
Observations imply that the predicted power spectrum of anisotropies
must be nearly independent of the scale on which the anisotropies are
probed, which implies that the inflationary phase responsible for
structure formation must be close to exponential acceleration. This is
true for slow-roll inflation, but also for the inflationary phase
obtained from the effective density once a non-zero scalar potential
is taken into account \cite{GenericInfl}. For more detailed comparisons
between theory and observations one needs to consider how
inhomogeneous fields evolve, which already requires us to relax the
strong symmetry assumption of homogeneity. The necessary methods are
not well-developed at the current stage (see
\cite{SphSymm,Horizon,SphSymmHam} for the basic
formulation), but preliminary calculations of implications on the
power spectrum have been undertaken nonetheless.
Ref.~\cite{PowerLoop} indicates that loop inflation can be
distinguished from simple inflaton models because the power depends
differently on scales. 

It turns out that this loop phase alone can provide a sufficient
amount of inflation only for unnatural choices of parameters (such as
extremely large $j$), and those cases are even ruled out by
observations already. At this point, the modified matter dynamics of
(\ref{effKG}) and its $\dot{\phi}$-term becomes important.
Classically, it is a friction term which is used for slow-roll
inflation. But in the modified regime at small scales the sign of the
term changes once $(a^{-3})_{\rm eff}$ is increasing. Thus, at those
small scales friction turns into antifriction and matter is driven up
its potential if it has a non-zero initial momentum (even a tiny one,
e.g., from quantum fluctuations). After the loop phase matter fields
slow down and roll back toward their minima, driving additional
inflation. The potentials need not be very special since structure
formation in the first phase and providing a large universe happen by
different mechanisms. When matter fields reach their minima they start
to oscillate and usual re-heating to obtain hot matter can commence.

Loop quantum cosmology thus presents a viable alternative to usual
inflaton models which is not in conflict with current observations
but can be verified or ruled out with future experiments such as the
Planck satellite. Its attractive feature is that it does not require
the introduction of an inflaton field with its special properties, but
provides a fundamental explanation of acceleration from quantum
gravity. This scenario is thus encouraging, but so far has not been
developed to the same extent as inflaton models.

\begin{figure}
\centerline{\includegraphics[width=17cm]{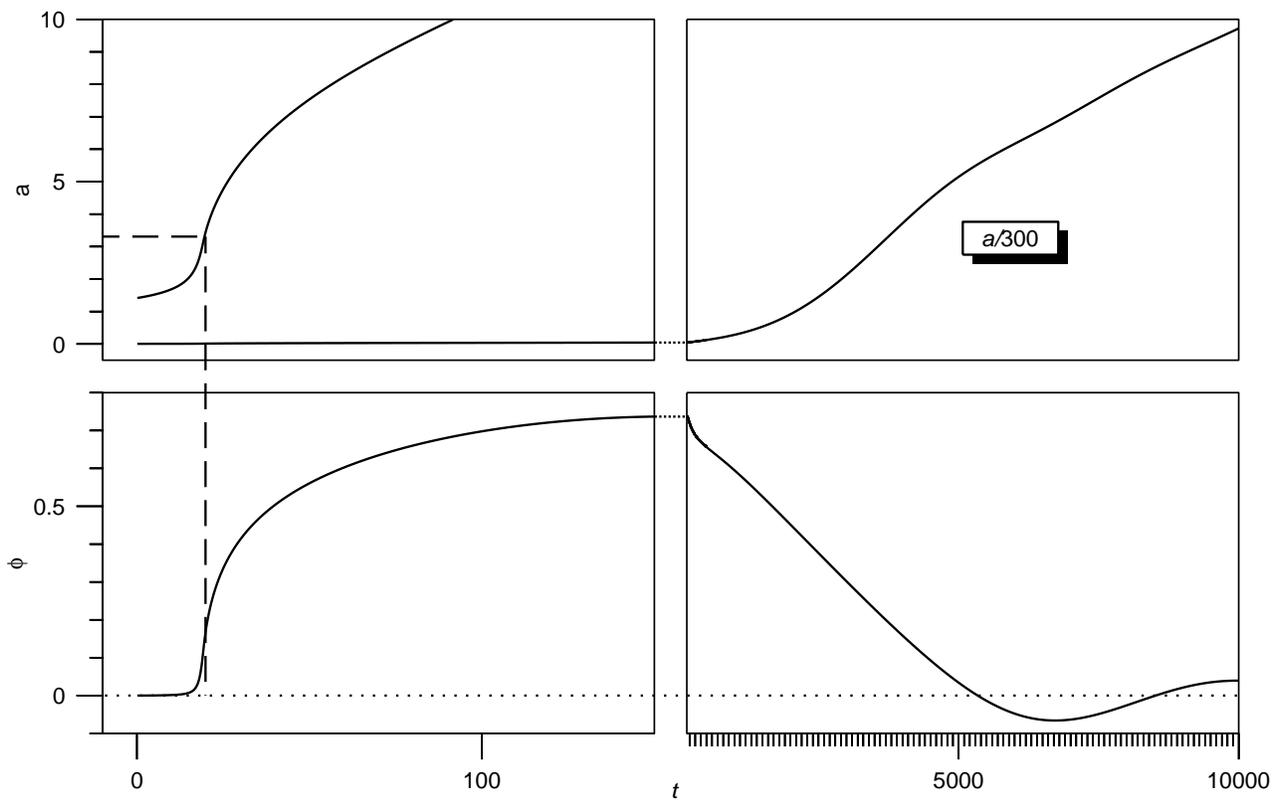}}
\caption{History of the scale factor (top) and inflaton (bottom) with
  the left hand side in slow motion. (Tics on the right horizontal
  axis mark increments in $t$ by 100.) The upper right data are
   rescaled so as to fit on the same plot. Units for $a$ and $\phi$
   are Planck units, and parameters are not necessarily realistic but
   chosen for plotting purposes. Dashed lines mark
  the time and scale factor where classical behavior of $(a^{-3})_{\rm
    eff}$ starts.
\label{Push}} 
\end{figure}

Even if we assume the presence of an inflaton field are its properties
less special than in the purely classical treatment. We still need to
assume a potential which is sufficiently flat, but there is now an
explanation of initial values far away from the minimum. For this we
again use the effective Klein--Gordon equation and the fact that
$\phi$ is driven up its potential. One can then check that for usual
inflaton potentials the value of typical initial conditions, as a
function of chosen ambiguity parameters and initial fluctuations of
the scalar, is just what one needs for sufficient inflation in a wide
range \cite{InflationWMAP,Robust}. After the modifications in the
density subside, the inflaton keeps moving up the potential from its
initial push, but is now slowed down by the friction term. Eventually,
it will stop and turn around, entering a slow roll phase in its
approach to the potential minimum.  Thus, the whole history of the
expansion is described by a consistent model as illustrated in
Fig.~\ref{Push}, not just the slow roll phase after the inflaton has
already obtained its large initial values.

One may think that such a second phase of slow-roll inflation washes
away potential quantum gravity effects from the early expansion. That
this is not necessarily the case has been shown in
\cite{InflationWMAP}, based on the fact that around the turning point
of the inflaton the slow-roll conditions are violated.  In this
scenario, structure we see today on the largest scales was created at
the earliest stages of the second inflationary phase since it was
enlarged by the full inflationary phase. If the second inflationary
regime did not last too long, these scales are just observable today
where in fact the observed loss of power can be taken as an indication
of early violations of slow-roll expansion. Thus, loop quantum
cosmology can provide an explanation, among others, for the
suppression of power on large scales.

There are diverse scenarios since different phases of inflation can be
combined, and eventually the right one has to be determined from
observations. One can also combine bounces and inflationary regimes in
order to obtain cyclic universes which eventually reach a long phase
of accelerated expansion \cite{InflOsc}. In particular, this allows the
conctruction of models which start close to a simple, static initial
state and, after a series of cycles, automatically reach values of the
scalar to start inflation. In this way, a semiclassical non-singular
inflationary model \cite{Emergent,Emergent2,EmergentLoop} is formulated
which evades the singularity theorem of \cite{InflSing}.

Current observations are already beginning to rule out
certain, very large values of the ambiguity parameter $j$ such that
from future data one can expect much tighter limits. In all
these scenarios the non-perturbative modification of the density is
important, which is a characteristic feature of loop quantum
cosmology. At larger scales above the peak there are also perturbative
corrections which imply small changes in the cosmological expansion
and the evolution of field modes. This has recently been
investigated \cite{PowerPert} with the conclusion that potential
effects on the power spectrum would be too small to be noticed by the
next generation satellites. The best candidates for observable effects
from quantum gravity thus remain the non-perturbative modifications in
effective densities.

\section{Conclusions and Outlook}

What we have described is a consistent picture of the universe which
is not only observationally viable but also mathematically
well-defined and non-singular. There are instances where quantum
gravity is essential, and others where it is helpful in achieving
important effects. The background independent quantization employed
here is very efficient: There are a few basic properties, such as the
discreteness of spatial geometry and the representation only of
exponentials of curvature, which are behind a variety of applications.
Throughout all the developments, those properties have been known to
be essential for mathematical consistency before they were recognized
as being responsible for physical phenomena.

For instance, for the singularity issue the basic properties are all
needed in the way they turn out to be realized. First, the theory had
to be based on densitized triad variables which now not only provides
us with the sign of orientation, and thus two sides of the classical
singularity, but also in more complicated models positions the
classical singularity in phase space such that it becomes accessible
by quantum evolution.  Then, the discreteness of spatial geometry
encoded in triad operators and the representation of exponentials of
curvature play together in the right way to remove divergences in
densities and extend the quantum evolution through the classical
singularity. These features allow general results about the absence of
singularities without any new or artificial ingredients, and lead to a
natural solution of a long-standing problem which has eluded previous
attempts for decades.  Symmetry assumptions are still important in
order to be able to perform the calculations, but they can now be
weakened considerably and are not responsible for physical
implications. The essential step is to base the symmetry reduction on
a candidate for full quantum gravity which is background independent
so as to allow studying quantum geometry purely.

Absence of singularities in this context is a rather general
statement about the possibility to extend a quantum wave function
through a regime which classically would appear as a singularity. More
explicit questions, such as what kind of new region one is evolving to
and whether it again becomes classical or retains traces of the
evolution through a quantum regime, depend on details of the relevant
constraint operators. This includes, for instance, quantization
ambiguities and the question whether a symmetric operator has to
be used. The latter aspect is also important for technical concepts
such as a physical inner product.

Here we discussed only isotropic models which are classically
described solely by the scale factor determining the size of space.
But a more realistic situation has to take into account also the shape
of space, and changes of the distribution of geometry and matter
between different points of space. The methods we used have been
extended to homogeneous models, allowing for anisotropic spaces, and
recently to some inhomogeneous ones, defined by spherical symmetry and
some forms of cylindrical symmetry. In all cases, essential aspects of
the general mechanism for removing classical singularities which has
first been seen only in the simple isotropic models are known to be
realized.\footnote{This does not refer to the boundedness of densities
  or curvature components for {\em all geometries}, which is known not
  to be present in anisotropic models \cite{HomCosmo,Spin} or even on
  some degenerate configurations in the full theory \cite{BoundFull}.
  What is relevant is the behavior on configurations seen along the
  dynamical evolution.}  Moreover, in the more complicated systems it
is acting much more non-trivially, again with the right ingredients
provided by the background independent quantization. Nevertheless,
since the inhomogeneous constraints are much more complicated to
analyze, absence of singularities for them has not yet been proven
completely.  The inhomogeneous systems now also allow access to black
hole and gravitational wave models such that their quantum geometry
can be studied, too.

Effective equations are a useful tool to study quantum effects in a
more familiar setting given by classical equations of motion. They
show diverse effects whose usefulness in cosmological phenomenology is
often surprising. Also here, the effects were known to occur from the
quantization and the transfer into effective classical equations,
before they turned out to be helpful. In addition to inflationary
scenarios and bounces which one can see in isotropic cosmologies,
modified densities have more implications in less symmetric models.
The anisotropic Bianchi IX model, for instance, is classically chaotic
which is assumed to play a role in the complicated approach to a
classical singularity \cite{BKL}. With the effective modifications the
dynamics changes and simplifies, removing the classical
chaos \cite{NonChaos}. This has implications for the effective approach
to a classical singularity and can provide a more consistent picture
of general singularities \cite{ChaosLQC}. Effective classical equations
can also be used to study the collapse of matter to a black hole, with
modifications in the development of classical singularities and
horizons \cite{Collapse}. This can now also be studied with
inhomogeneous quantum models which allow new applications for black
holes and cosmological phenomenology where the evolution of
inhomogeneities is of interest in the context of structure formation.

With these models there will be new effects not just in cosmology but
also for black holes and other systems which further check the overall
consistency of the theory. Moreover, a better understanding of
inhomogeneities evolving in a cosmological background will give us a
much better computational handle on signatures in the cosmic microwave
or even gravitational wave background, which may soon be testable with
a new generation of observations. One may wonder how it can be
possible to observe quantum gravity effects, given that the Planck
scale is so many orders of magnitude away from scales accessible by
today's experiments. The difference in scales, however, does not
preclude the observation of indirect effects even though direct
measurements on the discreteness scale are impossible, as illustrated
by a well-known example: Brownian motion allows to draw conclusions
about the atomic structure of matter and its size by observations on
much larger scales \cite{Brownian}. Similarly, cosmological
observations can carry information on quantum gravity effects which
otherwise would manifest themselves only at the Planck scale.

%\bibliographystyle{../preprint}
%\bibliography{../Bib/QuantGra}

\end{document}